\documentclass[conference, a4paper]{IEEEtran}
\IEEEoverridecommandlockouts
\usepackage{amsmath,amssymb,amsfonts}
\usepackage{algorithmic}
\usepackage{graphicx}
\usepackage{siunitx}
\usepackage{textcomp}
\usepackage{xcolor}
\usepackage{wrapfig}
\usepackage{tikz}
\usepackage{array} 
\usepackage{pgfplots}
\pgfplotsset{compat=1.17}
\usepackage[doi=false,isbn=false,url=false,eprint=false,backend=biber,style=ieee,]{biblatex}
\def\BibTeX{{\rm B\kern-.05em{\sc i\kern-.025em b}\kern-.08em
    T\kern-.1667em\lower.7ex\hbox{E}\kern-.125emX}}
\begin{document}

\title{An AI-based, Error-bounded Compression Scheme for High-frequency Power Quality Disturbance Data\\
}
\author{\IEEEauthorblockN{Markus Stroot\IEEEauthorrefmark{1}\IEEEauthorrefmark{2}, Stefan Seiler\IEEEauthorrefmark{1}, Philipp Lutat\IEEEauthorrefmark{1}, Andreas Ulbig\IEEEauthorrefmark{1}\IEEEauthorrefmark{2}}
\IEEEauthorblockA{\IEEEauthorrefmark{1}\textit{IAEW at RWTH Aachen University, Germany, e-mail: m.stroot@iaew.rwth-aachen.de}}
\IEEEauthorblockA{\IEEEauthorrefmark{2}\textit{Fraunhofer FIT, Germany, e-mail: markus.stroot@fit.fraunhofer.de}}}

\maketitle

\begin{abstract}
The implementation of modern monitoring systems for power quality disturbances have the potential to generate substantial amounts of data, reaching a point where transmission and storage of high-frequency measurements become impractical. This research paper addresses this challenge by presenting a new, AI-based data compression method. It is based on existing, multi-level compression schemes; however, it uses state-of-the-art technologies, such as autoencoders, to improve the performance. Furthermore, it solves the problem that such algorithms usually cannot ensure an error bound. The scheme is tested on synthetically generated power quality disturbance samples. The evaluation is performed using different metrics such as final compression rate and overhead size. Compression rates between 5 and 68 were achieved depending on the error bound and noise level. Additionally, the impact of the compression on the performance of subsequent algorithms is determined by applying a classification algorithm to the decompressed data. The classification accuracy only declined by 0.8--11.9 \%, depending on the chosen error bound.
\end{abstract}


\section{Introduction}
The contemporary power grid landscape is undergoing a continuous transformation, shifting away from large, synchronously-coupled, centralized generation units towards decentralized structures.
This transition significantly influences grid inertia, stability, and consequently, power quality \cite{li2022a}.
As a result, there is a growing importance in monitoring the quality of power supply systems to analyze and predict situations of quality disturbances.
However, the implementation of such monitoring systems generates substantial amounts of data, reaching a point where transmission and storage may become impractical. \cite{Parle2001,Wartmann2004}

Two avenues are possible at this point: Using edge-computing to process data directly where it is produced, thus eliminating the need for transmission and storage, or using efficient methods of data compression to reduce the amount of data that needs to be transmitted and saved. In this contribution the second avenue is explored, in order to provide a possible solution, when edge-computing might be impossible or undesirable to implement.

The topic of  power quality disturbance compression has been explored in literature in several ways. Santos et al. \cite{santos1997} use dyadic wavelet transform, achieving compression rates between 3 and 6, while maintaining a normalized mean squared error (NMSE) of below \(10^{-5}\). Several works have since build on this approach: A more modern version of the wavelet-based method was proposed by Ruiz et al. \cite{ruiz2021}. They implement several improvements on the original ideas, and claim compression rates of up to 510 , while maintaining NMSE rates of below \(3\times10^{-3}\). It should be noted however, that a sampling rate of \SI{200}{\kilo\hertz} was used, which is comparably high and therefore leads to more redundancy and higher compression rates.
A different path was taken by Wang et al. \cite{wang2019}. They employed compressed sensing (CS) \cite{donoho2006}, mainly to perform classification on the compressed signal. Their approach achieved a compression rate of 4 while claiming a NMSE of below \(2\times10^{-3}\). 

All the mentioned approaches are based on traditional signal processing techniques. However, modern data-driven techniques provide a new and promising way for data processing, which might be able to improve compression rates even further. In other areas, the application of concepts like autoencoders have started to be explored in order to leverage the power of machine learning for data compression \cite{liu2021,yildirim2018a}. Preliminary studies already show the possible benefits such techniques can bring in the field of power quality disturbance compression \cite{stroot2023}.
Therefore, we propose a new, AI-based compression algorithm. Its novelty originates in the use of AI-based compression schemes within a multi-stage algorithm in the field of power quality. Additionally, it introduces the concept of deterministic error-bounding in conjunction with AI-based compression.

First, section 2 explains the proposed method in detail. Second, section 3 gives an overview of the training and evaluation procedure for the AI-based approaches and the overall method. Next, section 4 presents the results of the evaluation procedure, showing the general performance and the classification study. Last, a short conclusion and outlook are given.

\section{Proposed Method}
\begin{figure}
    \centering
    \includegraphics[width=\columnwidth]{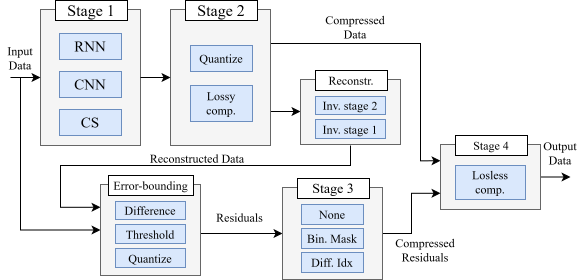}
    \caption{Proposed multi-stage compression method}
    \label{fig:comp}
\end{figure}

The proposed method is a multi-stage compression scheme shown in Fig. \ref{fig:comp}.
This structure is partly based on the SZ3 algorithm \cite{liang2023}; however, a domain specific evaluation of internal compression methods was performed and new ones introduced.

\subsection{First stage}
The first stage comprises one of the main contributions of this paper. 
Building upon \cite{stroot2023}, several data-driven compression schemes were implemented, trained, and tested.
Firstly, an efficient version of CS \cite{ravelomanantsoa2015} was applied as a more traditional candidate without the need for training resources.

Additionally, we tried two different forms of autoencoder.
An autoencoder is a machine learning concept often used for compression tasks. 
It works by combining two sub-structures: the encoder and the decoder.
They are trained back to back.
The encoder is forced by its structure to reduce the dimensionality of the input data, while the decoder reconstructs it.
The training loss (feedback)  is obtained from the difference of the original and the reconstructed signal. 
This way, both learn to minimize the resulting error on the given training data.
Afterwards they can be used separately to reduce and reconstruct data. \cite{goodfellow2016}

We focus on autoencoders based on Convolutional Neural Networks (CNN) and Recurrent Neural Networks (RNN).
Their differences and individual performance was already analyzed in \cite{stroot2023}.
This contribution aims to evaluate them further within a more realistic compression algorithm.

It should be noted that all algorithms considered in this stage provide no possibility of deterministically limiting the reconstruction error, although it is usually low for any appropriate application. Therefore, the reconstruction error of any following compression stage is also theoretically unbounded with respect to the original input.

\subsection{Second Stage}
In the second stage, a state-of-the-art lossy compression scheme is applied, to ensure maximum reduction of redundancy in the data. The lossy compression schemes tested are quantization, zfp \cite{lindstrom2014}, and  SZ3 \cite{liang2023}.

For quantization, a fixed interval and step size for all values is given and each value is then rounded to the closest step. Hence, only the respective step value needs to be saved for each data point. Quantization does not provide compression by itself, but it enables more efficient encoding and compression in subsequent steps. In this case, differential encoding is applied after quantization, which yields a data representation that is highly compressible by stage 4 lossless compression.

Zfp and SZ3 are state-of-the-art, generic floating point compression algorithms. Even though SZ3 is the algorithm on which this scheme is partially based, it is a lossy compression scheme and can be used as such in this context. Internally, several compression, reconstruction and error-encoding steps are performed; however, first stage compression in SZ3 is done using a linear predictor as opposed to CS or autoencoders.

It should be noted that all algorithms considered in this stage provide the possibility of deterministically limiting the reconstruction error. The error bound was chosen to be equal to the overall error bound of the scheme; however, it is thinkable that a more optimal internal bound might be determined by future investigations.

\subsection{Residuals and Third Stage}
One core problem with many lossy compression schemes such as autoencoders and CS is that there is no way to ensure a maximum reconstruction error. Similar to heuristics, it can be shown that they perform adequately in most cases; however, in edge cases or when confronted with data to far from their original design, no formal guarantees can be made. Therefore, the proposed method includes a strategy for introducing an upper bound for the reconstruction error.

This bound is enforced by performing the actual reconstruction directly after the compression and determining the point-wise difference of the original signal  \(\mathbf{x}\) and the reconstructed signal \(\mathbf{\overline{x}}\), the so-called residuals \(\mathbf{r}\), as shown in \eqref{eq:residuals}. All residuals that exceed the given error bound need to be appended to the compressed signal, so that they can be corrected after reconstruction, ensuring a maximum reconstruction error. 
\begin{align}
    \mathbf{r}&=\mathbf{x}-\mathbf{\overline{x}} \label{eq:residuals} \\
    \mathbf{\overline{x}}&=S_1^{-1}(S_2^{-1}(S_2(S_1(\mathbf{x})))) \\
    ||\mathbf{x}-(\mathbf{\overline{x}}+\mathbf{r}+\mathbf{e}_{quant})||_{\max}&\leq e_{bound} \label{eq:error_lim}
\end{align}

After calculation, the residuals are first masked, so all values below the error threshold are set to 0, since they do not need to be retained. The remaining residuals are quantized and differentially encoded in order to store them more efficiently. From \eqref{eq:residuals} and \eqref{eq:error_lim}  it becomes clear that this is viable as long as the quantization steps are chosen such that the error \(\mathbf{e}_{quant}\) is smaller the chosen error bound \(e_{bound}\). 

After quantization, three different approaches were tested to efficiently compress the masked and  quantized residuals. Firstly, the residuals were simply given to stage 4 lossless compression as is. Secondly, only non-zero residuals were saved together with their indices, so their association can be maintained. Thirdly, a binary mask is used to maintain the association.

The second technique, the indexing, saves all relevant residual values together with the index in the signal to which they should be applied. This is made more efficient using differential encoding. The third technique, binary masking, saves one bit per sample in the original signal. The bit can be either 0, meaning that no residual is applied, or 1, meaning that the next residual should be applied.

The second and third approaches are efficient as long as the number retained residuals is  low compared to the original signal size. As soon as this overhead of saving the indices or binary mask starts to outweigh the compression achieved on the complete residual vector, the benefit is lost. It is further expected that the overhead correlates with the selected error bound. A lower bound should lead to more retained residuals and, therefore, a lower compression rate and vice versa.

Which of these methods is more efficient depends on the sparseness of the retained residuals. As the number and precision of the residuals themselves is similar for both methods, only the size of the indexing data structure matters. The signals in this contribution are made up of 2560 samples. Therefore, the binary mask must be 2560 bit long. If we consider using 8-bit integer values for the differential indexing, we can easily determine that after \(\frac{2560}{8}=320\) retained residuals the binay mask becomes more efficient than the differential indexing. This corresponds to a sparsity of around \SI{87.5}{\percent}. 


\subsection{Fourth stage}
In this stage, last redundancies in the compressed signals and the retained residuals are eliminated. This has two rationals behind it. For one, there might be correlation between the residuals and the signals that can be exploited. Secondly, procedures like quantization of the signal or residuals makes the data look more regular, which enables lossless compression algorithms to further increase the compression rate.

Only lossless compression schemes are considered in this stage as to not introduce further compression error, which would negate the error-bounding procedure. In this case, bzip2 and gzip algorithms were evaluated, which are both general, state-of-the-art lossless compression schemes.

\section{Training and Evaluation Procedures}
The data-driven compression schemes were trained using synthetically generated power quality disturbance data according to the method described in \cite{stroot2023}. Briefly summarized, 15 common disturbance patterns were generated, 10 individual disturbances and 5 combined disturbances. Each signal is \SI{200}{\milli\second} long with a sampling frequency of \SI{12.8}{\kilo\hertz}, yielding \num{2560} samples per signal. Disturbances are generated using randomly generated parameters, such as disturbance amplitude and duration. Additionally, randomized disturbance ramp up and ramp down periods are introduced.

Hyperparameter tuning was performed for both convolutional and recurrent autoencoders according to \cite{stroot2023}. Both structures were tested using a compression rate of 8. A compression rate of 16 was used for the convolutional version as well, but discarded for the recurrent one due to poor convergence. The best performing parameter sets were fine tuned using a total of \num{10000} signals per disturbance class with a 90-10 training-evaluation split.

For reference, the CS approach was also included using compression rates of 8 and 16. A deterministic binary block diagonal matrix and a discrete cosine transform matrix were used as sampling and sensing matrices respectively. This compression method does not need to be trained, therefore it is simply constructed according to \cite{stroot2023} and applied in the proposed method.

For evaluation, the complete proposed compression method is applied to the set of \num{15000} evaluation signals. This allows the evaluation of several parameters. The main parameter is the final compression rate achieved depending on the chosen algorithms and error bounds. This was evaluated using noiseless data as well as in the presence of additive white Gaussian noise with a signal to noise ratio (SNR) of \SI{50}{\decibel} and \SI{40}{\decibel}. Further analysis is done on the amount of overhead introduced by the error-bounding relative to the compressed signal itself.

Lastly, this contribution aims to showcase, what kind of an impact an increased reconstruction error can have on other domain-specific algorithms that do follow up analyses on the reconstructed data. For this purpose, a neural network classifier was trained and evaluated on synthetically generated data. Afterwards, the classification is performed on the reconstructed evaluation data instead of the original data. By comparing the classification results, the impact of the reconstruction error on classification tasks can be shown.

\section{Results}
\subsection{General Performance}
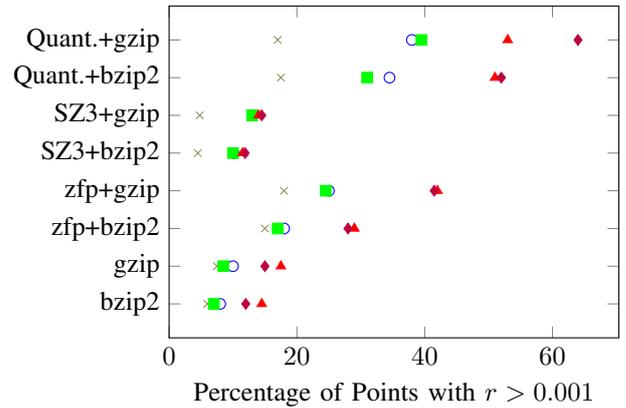
\begin{figure}
\centering
\begin{tikzpicture}
        \begin{axis}[
            scatter, only marks, scatter src=explicit symbolic,
            scatter/classes={
                a={mark=o,draw=blue,fill=blue},
                b={mark=x,draw=yellow!50!black},
                c={mark=square*,green},
                d={mark=triangle*,red},
                e={mark=diamond*,purple}},
            xmin=0,
            width=7.5cm,
            y=0.5cm,
            symbolic y coords={bzip2,gzip,zfp+bzip2,zfp+gzip,SZ3+bzip2,SZ3+gzip,Quant.+bzip2,Quant.+gzip},
            ytick=data,
            xlabel={Percentage of Points with $r>0.001$},
            bar width=0.4cm,
           enlarge y limits={abs=0.45cm},
            ]
            \addplot table[x=x,y=y,meta=label]{
                x        y                          label
                38      Quant.+gzip      a
                34.5      Quant.+bzip2    a
                13.1   SZ3+gzip           a
                10.1   SZ3+bzip2         a
                25      zfp+gzip           a
                18      zfp+bzip2          a
                10      gzip                    a
                8       bzip2                   a
                
                17      Quant.+gzip      b
                17.5      Quant.+bzip2    b
                4.8   SZ3+gzip           b
                4.5   SZ3+bzip2         b
                18      zfp+gzip            b
                15      zfp+bzip2          b
                7.5      gzip                    b
                6       bzip2                   b
                
                39.5      Quant.+gzip      c
                31      Quant.+bzip2    c
                13   SZ3+gzip           c
                10   SZ3+bzip2         c
                24.5      zfp+gzip            c
                17      zfp+bzip2          c
                8.5      gzip                    c
                7       bzip2                   c
                
                53      Quant.+gzip      d
                51      Quant.+bzip2    d
                14   SZ3+gzip           d
                11.5   SZ3+bzip2         d
                42      zfp+gzip            d
                29      zfp+bzip2          d
                17.5      gzip                  d
                14.5       bzip2               d
                
                64      Quant.+gzip      e
                52      Quant.+bzip2    e
                14.5   SZ3+gzip           e
                11.9   SZ3+bzip2         e
                41.5      zfp+gzip            e
                28      zfp+bzip2          e
                15      gzip                  e
                12       bzip2               e
            };
        \end{axis}
    \end{tikzpicture}
\caption{Comparison of compression rates for different algorithm combinations on power quality disturbance data}
\label{fig:combination-comp}
\end{figure}

Firstly, the compression rate of different algorithm combinations for stages 1, 2, and 4 was evaluated. The results can be seen in Fig. \ref{fig:combination-comp}.

Initial experiments with skipping stage 2 compression entirely and going directly to lossless compression turned out to be among the worst performing scenarios. Using the bzip2 algorithm even lead to worse compression rates than what the output of the first stage produced, meaning the algorithm added more overhead than it was able to remove. Gzip on its own performed marginally better. In combination with the stage 2 algorithms, gzip always performed better than bzip2 in these scenarios. Further investigations showed gzip performing better on smaller error bounds, while bzip2 performed better on larger bounds. Therefore, it was decided to apply both compression schemes and retain the smaller result of both. A single-bit flag is used to indicate to the decompression algorithm, which version was used.
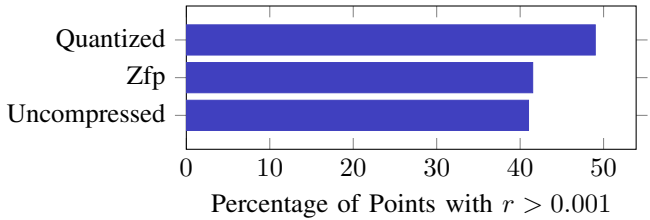
\begin{figure}
    \begin{tikzpicture}
        \begin{axis}[
            xbar, xmin=0,
            width=7.5cm,
            y=0.5cm,
            symbolic y coords={Uncompressed, Zfp, Quantized},
            ytick=data,
            xlabel={Percentage of Points with $r>0.001$},
            bar width=0.4cm,
           enlarge y limits={abs=0.45cm},
            ]
            \addplot+[fill=blue!50!gray,draw=blue!50!gray] coordinates {(41,Uncompressed) (41.5,Zfp) (49,Quantized)};
        \end{axis}
    \end{tikzpicture}
    \caption{Residuals larger than 0.001 after applying CS with a compression rate of 16 and either no further compression, zfp, or quantization}
    \label{fig:adderror-comp}
\end{figure}

Comparing the three options for stage 2, SZ3 achieved the lowest compression rates, being outperformed by zfp and quantization. This is partly due to the structure of the data. After stage 1 compression, only 320 or 160 samples remain per signal. As SZ3 usually performs much better on longer structures, it simply cannot use its full potential on this data.

At first glance, quantization seems the clear favorite, since it achieved the highest compression rates. However, looking at the additional residuals introduced by quantization and zfp versus the uncompressed data in Fig. \ref{fig:adderror-comp}, it is clear that zfp barely introduces any additional residuals, while quantization noticeably increases the number of residuals. A trade-off must therefore be considered between the higher compression rate of quantization and the higher error it introduces, leading to more overhead in terms of residuals. As SZ3 produces a similar error distribution to zfp, but much lower compression rates, it is from here on disregarded as an option for compression scheme. Similarly, RNN-based autoencoders are disregarded due to continued poor performance in all scenarios.

\begin{figure}
    \centering
    \begin{tikzpicture}[
  every axis/.style={ 
    ybar stacked,
    ymin=0,ymax=0.12,
    ylabel=relative size to uncompressed data,
    x tick label style={rotate=45,anchor=east,yshift=-0.25cm,xshift=0.25cm,font=\footnotesize},
    yticklabel style={
        /pgf/number format/fixed,
        /pgf/number format/precision=2
},
  legend style={at={(0.5,1)},anchor=north,yshift=-0.1cm,nodes=right,font=\footnotesize, legend columns=2,column sep=0.3cm},
    xtick=data,
    symbolic x coords={
      CAE8+Quant.,
      CAE8 + Zfp,
      CS8 + Quant.,
      CS8 + Zfp,
      CAE16 + Quant.,
      CAE16 + Zfp,
      CS16 + Quant.,
      CS16 + Zfp
    },
  bar width=4pt
  },
]

\begin{axis}[bar shift=-5pt,hide axis]
\addplot+[fill=blue!50!gray] coordinates
{(CAE8+Quant.,0.0253) (CAE8 + Zfp,0.04) (CS8 + Quant.,0.02333) (CS8 + Zfp,0.04133) (CAE16 + Quant.,0.01733) (CAE16 + Zfp,0.02333) (CS16 + Quant.,0.01266) (CS16 + Zfp,0.02333)}; 
\addplot+[fill=green!50!gray,draw=green!50!gray] coordinates
{(CAE8+Quant.,0.024) (CAE8 + Zfp,0.02133) (CS8 + Quant.,0.03666) (CS8 + Zfp,0.03466) (CAE16 + Quant.,0.05066) (CAE16 + Zfp,0.05) (CS16 + Quant.,0.05733) (CS16 + Zfp,0.056)}; \addlegendentry{Error Mask}
\end{axis}

\begin{axis}[hide axis]
\addplot+[fill=blue!50!gray] coordinates
{(CAE8+Quant.,0.0253) (CAE8 + Zfp,0.04) (CS8 + Quant.,0.02333)(CS8 + Zfp,0.04133) (CAE16 + Quant.,0.01733) (CAE16 + Zfp,0.023333) (CS16 + Quant.,0.01266) (CS16 + Zfp,0.02333)};
\addplot+[fill=yellow!50!gray,draw=yellow!50!gray] coordinates
{(CAE8+Quant.,0.012) (CAE8 + Zfp,0.00933) (CS8 + Quant.,0.01) (CS8 + Zfp,0.00866) (CAE16 + Quant.,0.01533) (CAE16 + Zfp,0.01533) (CS16 + Quant.,0.01133) (CS16 + Zfp,0.01)}; 
\addplot+[fill=red!50!gray,draw=red!50!gray] coordinates
{(CAE8+Quant.,0.01466) (CAE8 + Zfp,0.01333) (CS8 + Quant.,0.02866) (CS8 + Zfp,0.028) (CAE16 + Quant.,0.038) (CAE16 + Zfp,0.03733) (CS16 + Quant.,0.05) (CS16 + Zfp,0.05)}; 
\end{axis}

\begin{axis}[bar shift=5pt]
\addplot+[fill=blue!50!gray] coordinates
{(CAE8+Quant.,0.0253) (CAE8 + Zfp,0.04) (CS8 + Quant.,0.023333) (CS8 + Zfp,0.04133) (CAE16 + Quant.,0.01733) (CAE16 + Zfp,0.023333) (CS16 + Quant.,0.01266) (CS16 + Zfp,0.02333)}; \addlegendentry{Compressed Signal}
\addlegendimage{fill=green!50!gray,draw=green!50!gray}
\addlegendentry{Error mask}
\addlegendimage{fill=yellow!50!gray,draw=yellow!50!gray}
\addlegendentry{Binary mask}
\addplot+[fill=gray,draw=gray] coordinates
{(CAE8+Quant.,0.014) (CAE8 + Zfp,0.012) (CS8 + Quant.,0.012) (CS8 + Zfp,0.01) (CAE16 + Quant.,0.018) (CAE16 + Zfp,0.018) (CS16 + Quant.,0.012) (CS16 + Zfp,0.01133)}; \addlegendentry{Difference indices}
\addplot+[fill=red!50!gray,draw=red!50!gray] coordinates
{(CAE8+Quant.,0.01466) (CAE8 + Zfp,0.01333) (CS8 + Quant.,0.02866) (CS8 + Zfp,0.028) (CAE16 + Quant.,0.038) (CAE16 + Zfp,0.03733) (CS16 + Quant.,0.05) (CS16 + Zfp,0.05)}; \addlegendentry{Error}

\end{axis}
\end{tikzpicture}
    \caption{Comparison of average relative compressed size of  data, focusing on the overhead introduces by residuals}
    \label{fig:residuals-comp}
\end{figure}
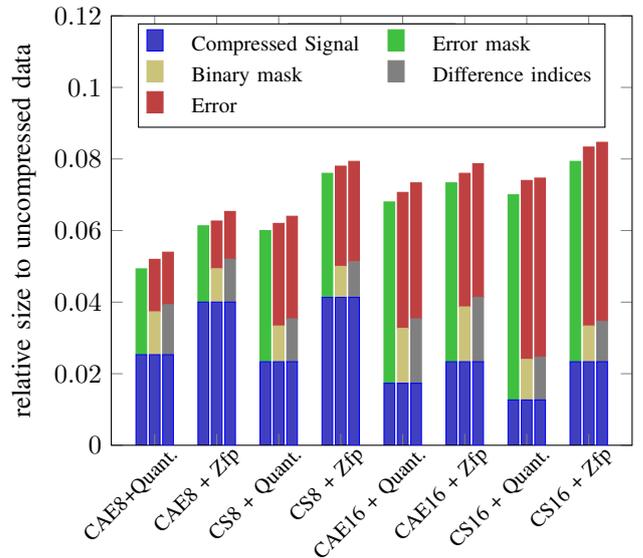

Looking more closely at the compression of the residuals in Fig. \ref{fig:residuals-comp}, it can be seen that simply handing the masked and quantized residuals to stage 4 compression yielded the most promising results, followed by the binary mask and then differential indexing. The reasoning here being the fact that the additional overhead in form of the indexing or bitmask structures is harder for the lossless algorithms to compress than the sparse residual vector on its own. Therefore, the direct compression of the masked and quantized residuals will be used in the following analyses.

\begin{figure*}
    \centering
    \includegraphics[width=\textwidth]{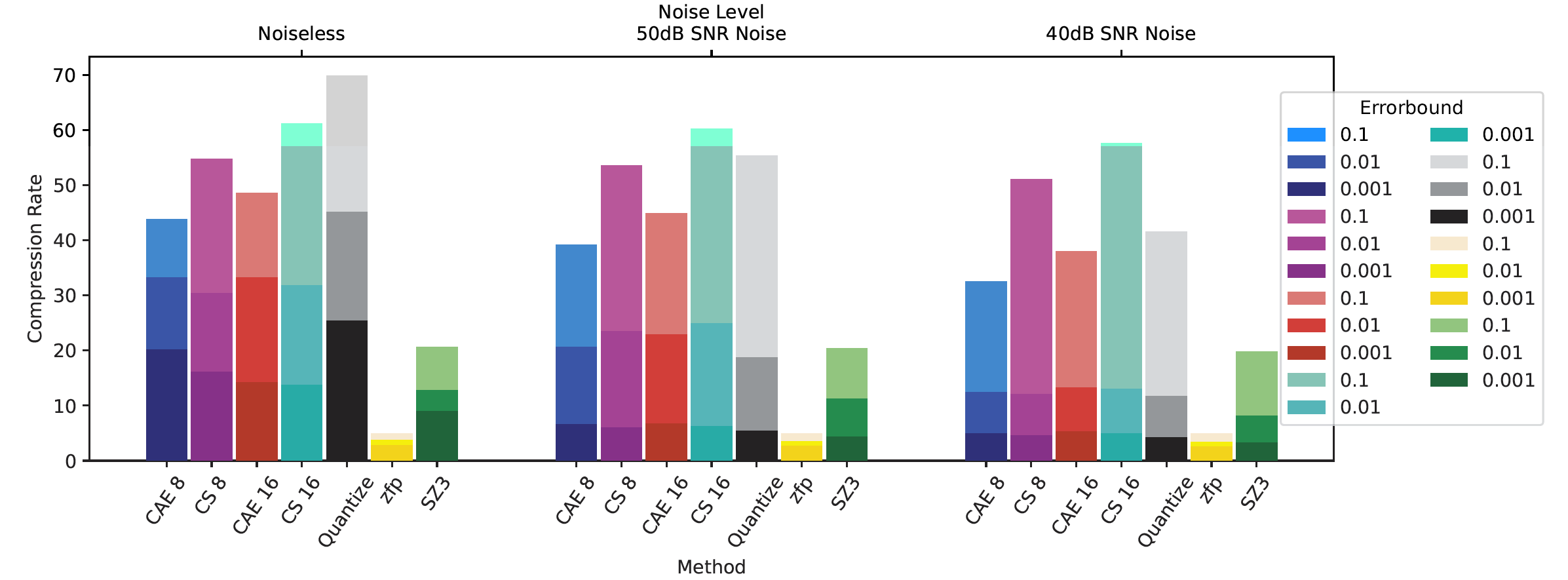}
    \caption{Comparison of the most efficient compression schemes for varying error bounds and noise level}
    \label{fig:comp-all}
\end{figure*}

The results in Fig. \ref{fig:residuals-comp} already provide an initial glace at the overall performance of  the complete compression scheme using different stage 1 and 2 compression schemes. A more complete overview can be seen in Fig. \ref{fig:comp-all}. Several aspects can be observed:

First, using state-of-the-art compression schemes such as zfp or SZ3 on the data directly always yielded the lowest compression rates, showing the possible gain by using specialized compression schemes.

Second, using simple quantization without any stage 1 compression on noiseless data performed best for all demonstrated error bounds. However, as soon as noise is introduced, which is to be expected is real application scenarios, the performance drops significantly. In the \SI{50}{\decibel} SNR scenario, methods based on CS and autoencoders usually outperform quantization, especially for lower error bounds.

Third, for higher errors bounds, such as 0.1, the CS-based approaches yielded the best results, while for lower error bounds, like 0.001, usually the autoencoder is slightly ahead. This is most probably due to the denoising effect of CS, which makes the data easier to compress by following stages, but also increases the reconstruction error slightly. For higher error bounds, the additional overhead from residuals is not as impactful, while for lower thresholds, the more faithful representation of the autoencoder succeeds due to fewer residuals.

The best performing approach depends on the chosen error bound and noise level. While for higher bounds CS comes out ahead, lower bounds tend to perform better with autoencoders. It should be noted, that the autoencoders were trained using noiseless data. Using noisy data in the training procedure has been shown to increase their performance even further in preliminary tests.

\subsection{Classification study}
In order to showcase the impact of the compression on an actual application of the data, a classification study has been performed. For this, a simple convolutional neural network was trained for classification of the 15 power quality disturbance classes used in this contribution \cite{mohan2017}. After training the classifier achieved an accuracy of \SI{98}{\percent} . Hence, the classifier is capable of identifying the signals with a high degree of accuracy. Most of the wrong classifications tend to be of signals that are ambiguous even to the human observer. Confusion between the disturbance \textit{sag} and combined disturbance \textit{flicker sag} for example are likely, as they tend to be nearly indistinguishable depending on fault amplitude and duration.

\begin{figure}
    \centering
    \includegraphics[width=\columnwidth]{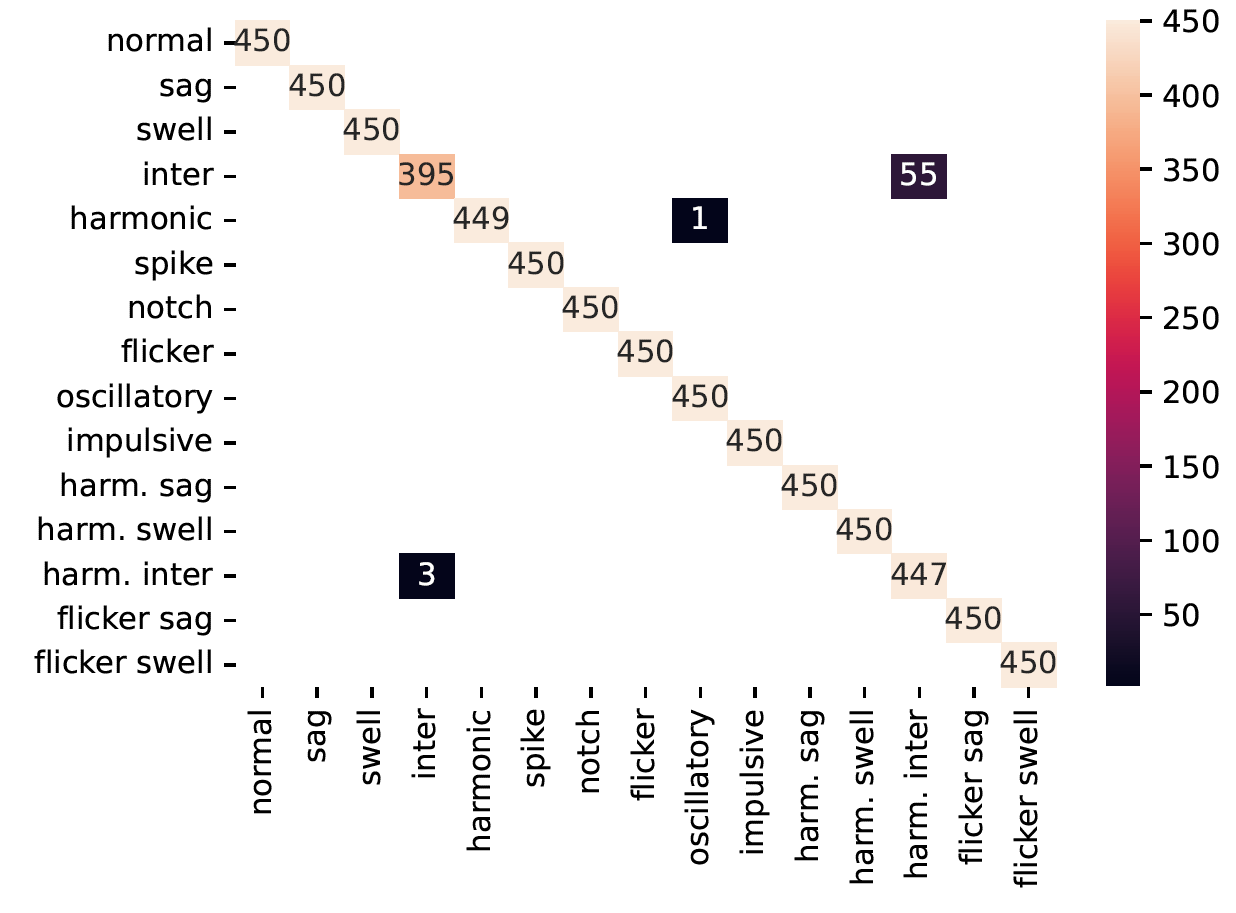}
    \caption{Confusion matrix of the trained classifier for the data compressed using a convolutional autoencoder scheme and an error bound of 0.001}
    \label{fig:confusion-001}
\end{figure}

\begin{figure}
    \centering
    \includegraphics[width=\columnwidth]{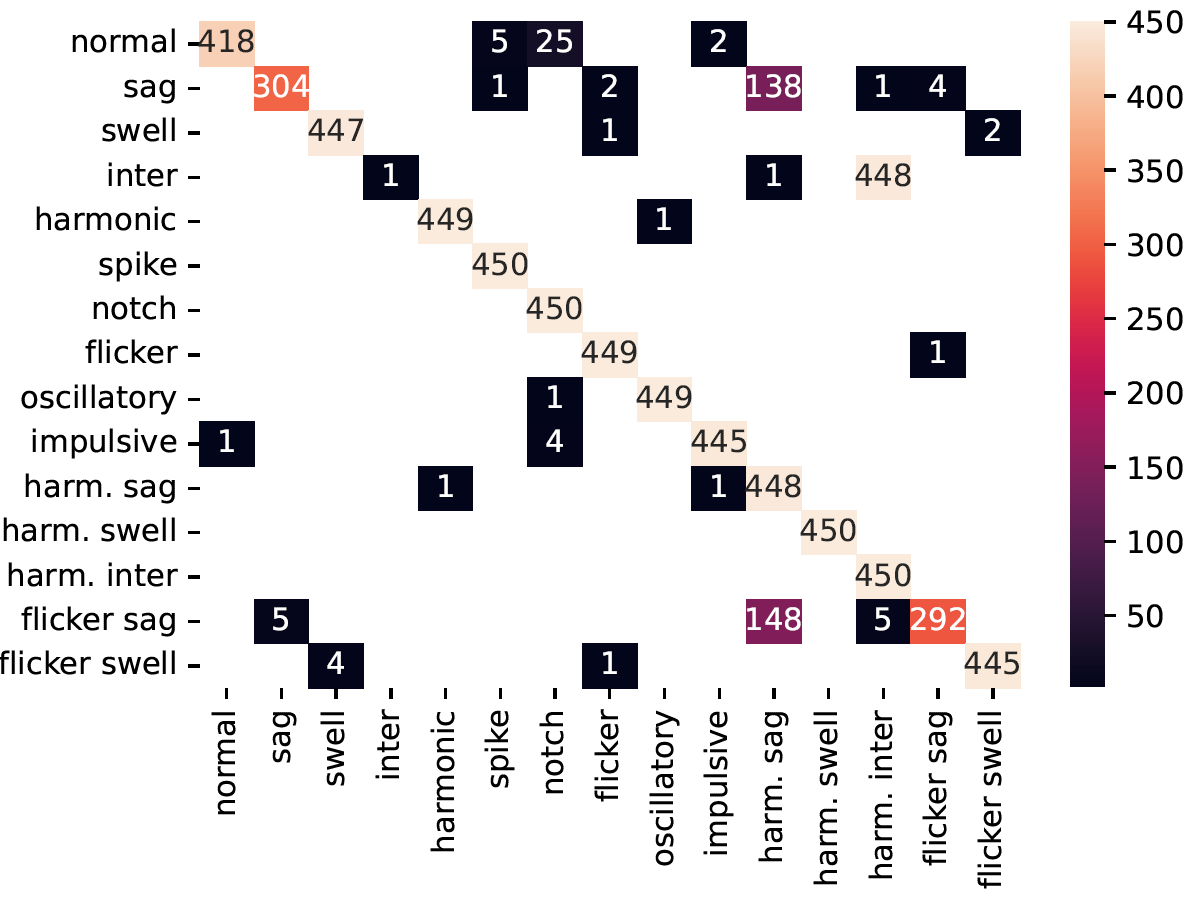}
    \caption{Confusion matrix of the trained classifier for the data compressed using a convolutional autoencoder scheme and an error bound of 0.01}
    \label{fig:confusion-01}
\end{figure}

To reduce any confusion introduced by the wrongly classified signals, they were eliminated from the pool. To ensure an even distribution of all classes, the pool was then reduced to 450 signals per disturbance by randomly eliminating signals. The classifier should now be able to correctly identify all signals in the pool. Now, different versions of the compression scheme were applied to the data. Any new misclassifications can therefore be considered due to the reconstruction error of the compression scheme. The results in Fig. \ref{fig:confusion-001} and \ref{fig:confusion-01} show the compression based on the convolutional autoencoder and error bounds of 0.001 and 0.01 respectively. For the lower bound, the classification only has minor problems distinguishing interruptions and harmonic interruptions, which is reasonable. At a higher error bound, much more misclassifications are present, especially for sags, flicker sags, and harmonic sags. Compressed sensing and quantization-based approaches yielded similar results, leading to the conclusion that the error bound itself is more relevant to the classification than the underlying compression algorithm.
\section{Conclusion}
In this contribution, a new algorithm for power quality disturbance data compression based on modern, data-driven approaches was introduced. The proposed algorithm outperforms state-of-the-art, general compression schemes significantly in terms of compression rate, while providing similar error-bounding capabilities. In comparison with other advances from literature, the most advanced examples reach similar compression rates; however, they do not deterministically ensure any error bound, but only provide maximum reconstruction errors from exemplary analyses. Further, it could be shown that depending on the chosen error threshold, subsequent processing such as classification is still possible with only minor impact of the reconstruction error.

However, it should be noted, that similarly high compression rates for the same error bounds were achieved using simple quantization in combination with lossless compression and retention of residuals. It stands to reason that a more simple approach such as this could be sufficient as well, especially in environments with limited computational resources. On the other hand, the more complex, data-driven approaches have been shown to be more robust towards interference by noisy data. An overview of advantages and disadvantages of the proposed scheme is given in Tab. \ref{tab:adv}.

\begin{table}[tb]
    \centering
    \caption{Advantages and disadvantages of the proposed scheme}
    \label{tab:adv}
    \begin{tabular}{>{\centering\arraybackslash}m{0.45\linewidth}>{\centering\arraybackslash}m{0.45\linewidth}} \hline
         \textbf{Advantages}  &  \textbf{Disadvanatges}\\\hline 
         \begin{itemize}
             \item high compression rates
             \item deterministic error bounding
             \item robust against noise
         \end{itemize} & \begin{itemize}
             \item training effort
             \item dependence on training data
             \item higher computational effort during compression
         \end{itemize}\\\hline 
     \\ \end{tabular}
\end{table}

In order to determine which approach is more suitable in general, or in specific cases, more research is necessary. One topic of future research is the impact of different compression schemes on other data processing tasks for power quality disturbance data. Additionally, more insight can be gained into the sensitivity of different classification schemes on the accuracy of the input data. Finally, more fine-tuning of the proposed compression scheme is always possible. Here, the training of autoencoders with noisy data already shows signs of benefit. A more complex training method, where the neural network is fine-tuned within the complete compression scheme is also possible. To support further research, the training data has been made available publicly.\footnote{DOI: 10.5281/zenodo.11393807 }

\section*{Acknowledgment}
\begin{wrapfigure}{r}{0.13\textwidth}
    \vspace{-\baselineskip}
    \vspace{-\baselineskip}
  \includegraphics[width=0.12\textwidth]{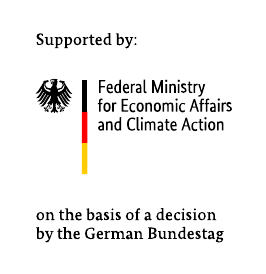}
  \vspace{-\baselineskip}
  \vspace{-\baselineskip}
\end{wrapfigure}
This project received funding from the German Federal Ministry for Economic Affairs and Climate Action under the agreement no. 03EI4048Q (Quirinus Control).

\printbibliography

\end{document}